# Sub-micromolar imaging of intrinsic chromophores by two-photon photothermal microscopy captures mitochondrial response to chemotherapy

Nathaniel Hai[1,3], Chinmayee Vallabh Prabhu Dessai[2,3], Dingcheng Sun[2,3], Jianpeng Ao[1,3], Pin-Tian Lyu[1,3], Yifan Zhu[1,3], and Ji-Xin Cheng[1,2,3]*

[1]Department of Electrical and Computer Engineering, Boston University, Boston, MA, 02215

[2]Department of Biomedical Engineering, Boston University, Boston, MA, 02215

[3]Photonics Center, Boston University, Boston, MA, 02215

*Corresponding author, email: jxcheng@bu.edu

**Abstract**

Intracellular chromophores (e.g., NADH and FAD) play a central role in regulation of cellular metabolism. Though autofluorescence has been extensively used for label-free mapping of chromophores inside a cell, its sensitivity and molecular specificity are constrained by the low quantum yield and the fluorescence spectral overlap. Here, we address these challenges by employing a photothermal approach to measure the optical absorption of chromophores rather than its autofluorescence. By combining near-infrared pump and visible probe beams, our two-photon photothermal (2PPT) microscope exploits localized thermal transients generated through two-photon absorption, enabling detection of chromophore-specific signatures beyond the reach of autofluorescence. We demonstrate sub-micromolar limit of detection for the metabolic coenzymes NADH and FAD of 0.87 μM and 0.99 μM, respectively. Such high sensitivity enables differentiating the influence of different mitochondria shapes on metabolism activity. Importantly, the fluorescence crosstalk-free 2PPT can identify the biomolecular source of contrast from cellular mitochondria in a label-free manner based on spectroscopy. 2PPT microscopy is utilized to study metabolic alterations of mitochondria in cancer under chemotherapy at the single organelle level.

# MAIN TEXT

## Introduction

Label-free, high-resolution imaging has been the foundation of sensing metabolism-related intrinsic chromophores in their native environment with minimal perturbation to their function (1,2). Building on early works showing applicability of intrinsic fluorescence (i.e., autofluorescence) to serve as a cancer biomarker (3,4), the field has rapidly accelerated with subsequent studies on its microstructural and functional mapping (5–8). A consensus guidelines aimed to standardize data collection across researchers in the field of label-free metabolic imaging was recently published (9). Meanwhile, by utilizing multiphoton excitation for better spatial confinement and spectral selectivity (10), the optical redox ratio derived from autofluorescence of NADH and FAD was shown to correlate with biochemical based assays that quantify the cellular phosphorylation activity and redox state (11). Harnessing fluorescence lifetime microscopy (FLIM) together with the emission intensity to measure the collective optical metabolic index has been identified as a more robust indicator of metabolic response to drug action at a cellular level, owing the distinct decay times of bound and unbound states of the biomolecule (12). Noninvasive quantification of metabolic activity based on multiparametric imaging further identified alterations in specific metabolic pathways and characterized the response of a single cell to them (13,14). Adopting intrinsic chromophore fluorescence as a contrast agent, FLIM techniques were applied for in-vivo characterization of biological tissues (15). Since then, intravital imaging modalities were carefully designed to detect metabolic activity of cells within the tumor microenvironment potentially enable in-vivo redox state mapping (16–20). More recently, studies on patient-extracted cellular organoids and tissues have emerged (21–23). Collectively, these works highlight the growing interest in such label-free noninvasive imaging approaches to study subcellular metabolic activity and its potential to serve as a diagnosis and therapeutic biomarker.

While the above-mentioned autofluorescence-based modalities have been widely used for sensing metabolism-relevant biomolecules in tens of micro molars, they are characterized by inherently weak signal levels (24) that are largely influenced by the properties of surrounding media like temperature, pH and binding partners (25). This native fluorescence originates from radiative decay of the molecule after absorption of excitation light, which is a less probable de-excitation pathway compared to nonradiative pathways (26) with a modest quantum yield of 2-10% for metabolic coenzymes such as NADH and FAD (25,27). Several research groups have addressed the low efficiency issue by creative designs of nanoprobes introduced into the medium of interest and detected by optical absorption or scattering mechanisms (28–30), achieving impressive limit of detection (LOD) in the sub-micro molar regions (31). However, the injection of exogenous probes can modify the natural biomolecular behaviors of the metabolism reporters rendering these approaches ill-posed for highly sensitive detection of the metabolic coenzymes in biomedically relevant scenarios.

Targeting the generated heat during light absorption by the target chromophores has potential to alleviate the above-mentioned challenges. Upon absorption the chromophore enters an excited state, from which relaxation back to the ground state often occurs through radiative decay (fluorescence emission) or non-radiative decay (e.g., internal conversion, vibrational relaxation) that contributes to heat. In low quantum efficiency biomolecules, energy transfer mechanisms are the second-most dominant after pathways that generate local heat (32). This implies that efficient sensing of these biomolecules could, in principle, be done via sensing the photothermal effect with extreme signal boost between 10-20 times compared with the autofluorescence detection. Indeed, previous efforts to optically sense the generated heat after photo-absorption has reached single chromophore sensitivity in a synthetic DNA construct (33). Furthermore, photothermal imaging within bio-compartments have achieved label-free detection of specific chromophores including hemoglobin and nanoparticles (34–36). Notably, photothermal imaging modalities that rely on multiphoton excitation benefit from several advantages owing the integration of nonlinear excitation with subsequent photothermal sensing and have shown the capacity to sense cytochrome molecules in kidney cells (37).

In the present study, we investigate 2 photon photothermal (2PPT) detection of metabolic coenzymes based on a comparison with standard autofluorescence approaches. We test the potential of imaging the photothermally excited biomolecules to deliver higher sensitivity and specificity given fixed experimental conditions. Specifically, 2PPT utilizes the highly efficient heat generation by the intrinsic chromophores compared to their fluorescence signal generation, delivering up to 20 times better LOD of these biomedically relevant coenzymes in low-concentration environment. Concurrently, 2PPT is shown to differentiate the influence of energetic perturbations in various mitochondria morphologies, which suggests how conformational changes might be associated with the organelle's metabolic activity. In addition, our approach delineates modifications in metabolic activity of live cells and tumor spheroids after targeted energetic perturbations and after chemotherapy drug treatment, respectively.

## Results
### A two-photon photothermal (2PPT) microscope

**Figure 1(A)** illustrates the two-photon photothermal microscope (2PPT), which utilizes a pump beam (red) for excitation and a probe beam (green) for label-free imaging of the photothermally generated heat by the target biomolecules in the sample. Our implementation employs a dual-port solid-state ultrafast laser oscillator as the light source for both beams, whereas the tunable near-infrared (NIR) port is used for excitation and combined with the frequency doubled output of the fixed 1045 nm port for heat sensing with a visible wavelength (522.5 nm). The generation of 2PPT signal can be described as a three-step process and is illustrated in **Figure 1(B)**. After light absorption by the target biomolecule, a relatively fast (~picoseconds) thermal relaxation within the excited electronic level occurs, after which the molecule can undergo a slower decay (nano- to micro-seconds) to a high vibrational state within the ground state and emit a photon (26). This process has low quantum yield of 8-10% for FAD and even lower efficiency of 2-4% for NADH (25, 27). However, the more probable event for the electronically excited molecule is to undergo a slower decay (tenths to tens of micro-seconds) back to the ground state without emitting any photon. This process generates a locally distributed transient heat that changes the local refractive index (38,39). Thus, the expression of a specific biomolecule, namely a chromophore, can be reliably measured through optically sensing the change in local refractive index with high sensitivity. To put this hypothesis to test, we use the described pump-probe configuration to sense the locally generated heat within sterile solutions of metabolic coenzymes, namely NADH and FAD, after two-photon absorption. As illustrated in **Figure 1(A)**, after the combination of the two beams via the dichroic mirror, they are relayed toward the sample plane after passing through a pair of galvanometric mirrors in order to laser scan the two beams at the back focal plane of a 60x 1.2 NA water immersion microscope objective (MO). After interacting with the sample, the two beams are collected by a 1.4 NA oil-immersion condenser lens (CL) equipped with a variable aperture. After that, the excitation beam is blocked by the chromatic filter and only the probe beam makes its way toward the single pixel photodetector connected to a digitization system based on lock-in amplifier (LIA) detection (see Materials and Methods for further details). As mentioned above, the timescale of the chromophore's non-radiative decay is in the range 0.1-10 µs, which also sets the timescale for the transient heat generation that serves as the contrast mechanism in 2PPT microscopy. Thus, to optimize the signal collection, we used an external acousto-optic modulator (AOM) to control the heat generation process. The AOM modulation was set to $f$=125 KHz, which mitigates the $1/f$ laser noise dominant at low frequency (<100 KHz) while satisfying the requirement for photothermal signal detection (100 KHz< $f$ <10,000 KHz). Decreasing the modulation frequency beyond that point shows minimal amplitude increase in signal and more prominent noise increase, which overall reduces the signal-to-noise ratio (SNR). Inset plots in **Figure 1(A)** show the signal time traces of pump beam (red plot) and the result in waveform of the probe beam (green plot). One can observe that the initially un-modulated probe beam takes the shape of rapid local heating followed by a gradual heat dissipation exactly at the pump beam modulation frequency, as expected by the heat propagation equation (39). Notably, the photothermal signal between consecutive pulse trains decays back to zero, which implies the complete dissipation of accumulated heat from the impacted area that is important for optimal signal extraction. The modulation parameters for our experiments, e.g., the duty cycle, modulation frequency, and decay time were verified based on simulation of the heat generation process of NADH solution (see Materials and Methods and Figure S1). An image is formed by two-dimensional scan of the pump and probe beams together via the galvo mirrors.

After we validated that two-photon absorption occurs in the sample, both by inherent optical sectioning and by power law scaling (Figure S2) (40,41), we tested the 2PPT sensing performance for metabolic coenzymes. **Figure 1(C)** illustrates the 2PPT signal in the case of NADH and FAD 0.25 mM solutions when using pump excitation at 720 nm and 780 nm. Our results closely match the two-photon action cross sections for the two biomolecules (24,40), emphasizing that signal generation in both methods is equivalent up to a factor that favors 2PPT (Supplementary text). We further tested the sensitivity of 2PPT in detecting the coenzymes and compared it with that of conventional two-photon autofluorescence (2PAF). **Figure 1(D)** and **Figure 1(E)** plot the observed signal for different concentrations of NADH solutions obtained with 2PPT and 2PAF, respectively, along with a linear fit from which the LOD for each method can be calculated based on the residuals standard deviation and the slope (see Materials and Methods and Supplementary text) (42). The 2PPT data points show lower residuals relative to the linear fit [**Figure 1(D)**] across the measured NADH concentrations (0.98-125 µM) compared with the plot obtained with 2PAF [**Figure 1(E)**]. Importantly, when we used these curves' linear fit to calculate the LOD for NADH, we found that 2PPT is almost 20 times more sensitive to the NADH concentration than 2PAF (0.87 µM versus 17.5 µM). Remarkably, this value agrees well with the 2-4% quantum efficiency of NADH (25, 27), considering the rest of excitation photons mostly contribute to heat generation. Using similar approach, we further verified our assumptions about the 2PPT order of magnitude higher sensitivity compared to 2PAF in the case of FAD. **Figure 1(F)** and **Figure 1(G)** plot the 2PPT and 2PAF signals obtained for different concentrations of FAD solutions and the best linear fit. Also equivalent to the 8-10% quantum efficiency of FAD, we found that 2PPT gives an order of magnitude better sensitivity to the concentration of FAD when compared with 2PAF (0.99 µM versus 11.6 µM). While it shows lower sensitivity to FAD compared with NADH, which is expected based on our assumption of lower autofluorescence emission-higher photothermal signal (and vice-versa), 2PPT still exhibits LOD below the micro-molar concentration that is better than fluorescence-based approaches. Note that the improved sensitivity of 2PAF to sense FAD (compared with NADH) is implicated by the tighter statistics it shows (**Figure 2(E)** versus **Figure 2(G)**). Overall, the high detection sensitivity of 2PPT to both NADH and FAD makes it a good candidate to study metabolism reprogramming in cancer cells and spheroids, where sensitive detection can image subtle alterations in the energetic balance after treatment.

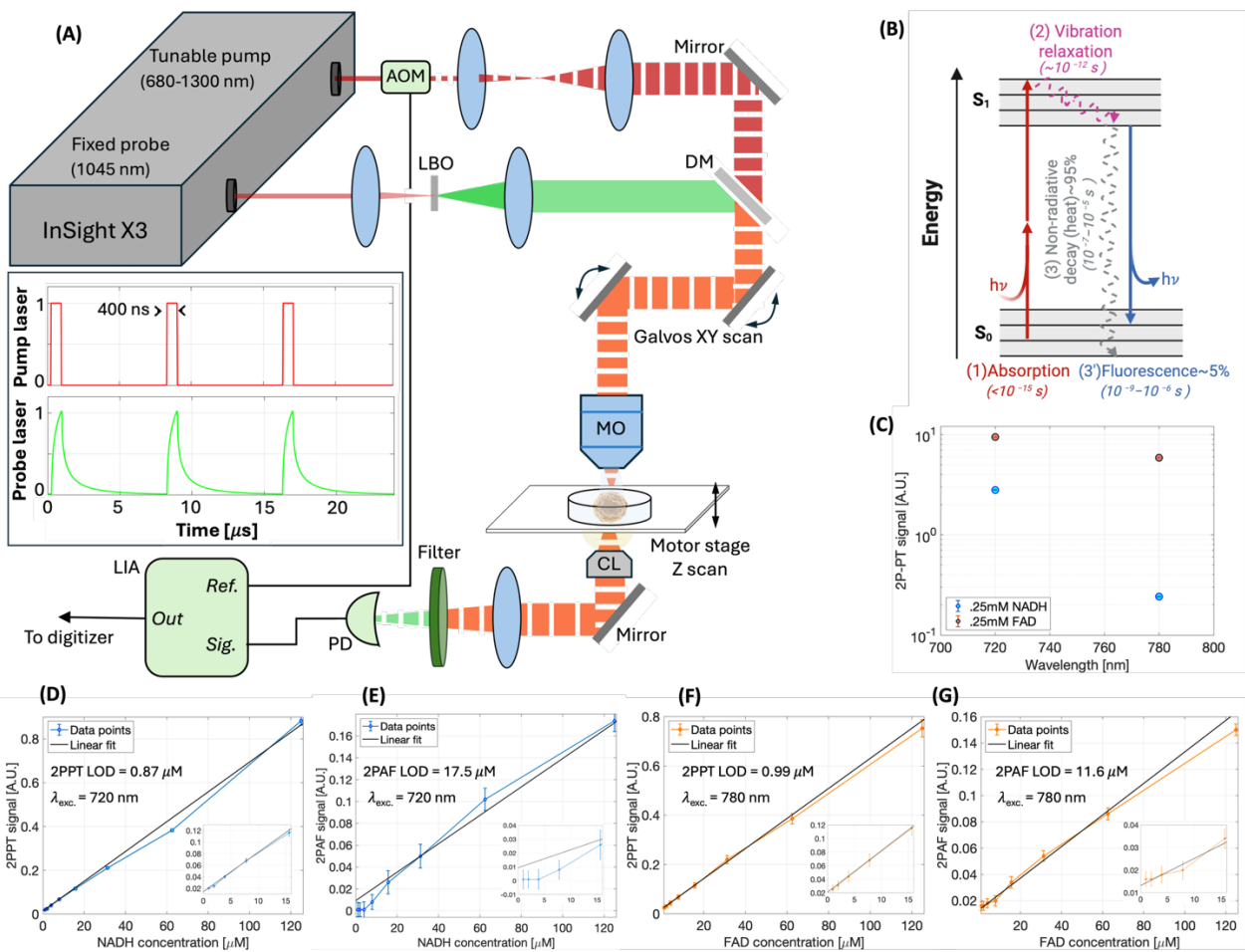

**Figure 1. Two-photon photothermal (2PPT) microscopy working principle and limit of detection.** (A) 2PPT microscope benchtop implementation. Inset shows time trace of pump laser (red) and probe laser (green). AOM: acousto-optics modulator; LBO: Lithium Triborate crystal; DM: dichroic mirror; MO: microscope objective; CL: condenser lens; PD: photodetector; LIA: lock-in amplifier. (B) Electronic and vibration energy bands diagram illustrating the three-step process of signal generation in 2PPT and 2PAF along with characteristic time scales (created with BioRender; www.biorender.com). (C) 2PPT signal of NADH and FAD sterile samples for estimating their action cross section. (D), (E) Plots of 2PPT and 2PAF signal as a function of NADH concentration and the linear fit from which limit of detection is calculated. Excitation wavelength is 720 nm. (F), (G) Plots of 2PPT and 2PAF signal as a function of FAD concentration and the linear fit from which limit of detection is calculated. Excitation wavelength is 780 nm. Inset of (D)-(G) shows data points and linear fit in low concentrations range.

**2PPT microscopy senses metabolism coenzymes with enhanced specificity**

In the next experiment we characterize the imaging capability of 2PPT to dynamically report the concentration and localization of metabolic coenzymes and compare it with the widely used 2PAF modality. We use cultured SK-OV-3 cancer cells, grown in a 2D layer and freshly imaged within 45 minutes out of incubation in 37°C (see Materials and Methods). **Figure 2(A)** and **Figure 2(D)** illustrate the micrographs of cells from the same field of view (FoV) obtained using 2PPT and 2PAF, respectively. The excitation wavelength is set to $\lambda_{pump}$=720 nm to target NADH molecules expression from subcellular organelles. For 2PAF detection, suitable emission filters were placed in front of the fluorescence detector according to the autofluorescence emission spectrum of the coenzyme (43). For more details about the 2PAF imaging setup see Materials and Methods. Appreciable signal comes from cytoplasmic mitochondria, and the characteristic tubular-shaped mitochondria are observed along with the oval-shaped mitochondria. **Figure 2(B)** depicts a representative single cell from the imaged FoV, where both mitochondria shapes can be observed clearly in the surrounding of the cell nucleus and some are marked with white (tubular) and yellow (oval) arrows. While the tubular, rod shape is widely associated with mitochondria structure, the oval, blob shape is also a common conformation of mitochondria that is associated with aging processes in cells, under-stress and diseased cells (44–46). Intensity histograms of selected cellular region (blue bins) and background region (orange bins) are given in **Figure 2(C)**, from which signal to background ratio (SBR) of 14.27 dB for 2PPT was calculated based on the means approach. **Figures 2(D)-(F)** are the 2PAF equivalents of **Figures 2(A)-(C)** and describe metrics obtained for the same cells FoV. Notably, 2PAF shows correspondence with the intense oval-shaped mitochondria observed in 2PPT, however the NADH signal associated with the tubular mitochondria seems highly scattered limiting the visibility of a single mitochondrion, which could arise from lack of sensitivity for the specific biomolecule. Indeed, comparison of the intensity histograms for both modalities [**Figure 2(C)** versus **Figure 2(F)**] shows more than 4 times higher SBR for 2PPT compared with 2PAF. Since the expression of cellular NADH is mainly associated with energy production and mostly reside in cell mitochondria, it is crucial that as a cellular metabolism imaging system, 2PPT have the capability to delineate the difference between the two conformations of mitochondria. While elongated shape mitochondria promote ATP production and can be seen in healthy, energy-demanding cells, the oval/blob shaped are the fragmented version of the mitochondria commonly observed during cell stress or death processes (47). To validate the observed structure of mitochondria in the tested cell line, we labeled fresh cells with Mito tracker and obtained micrographs using two-photon excited fluorescence (2PEF) imaging. Figure S3 illustrates different regions from the same dish of labeled cells, which confirms both types of mitochondria shapes exist in the tested cell line. Overall, the described tests and comparison show that 2PPT can detect NADH signal from single mitochondrion, regardless of its shape, with better sensitivity than the equivalent 2PAF approach.

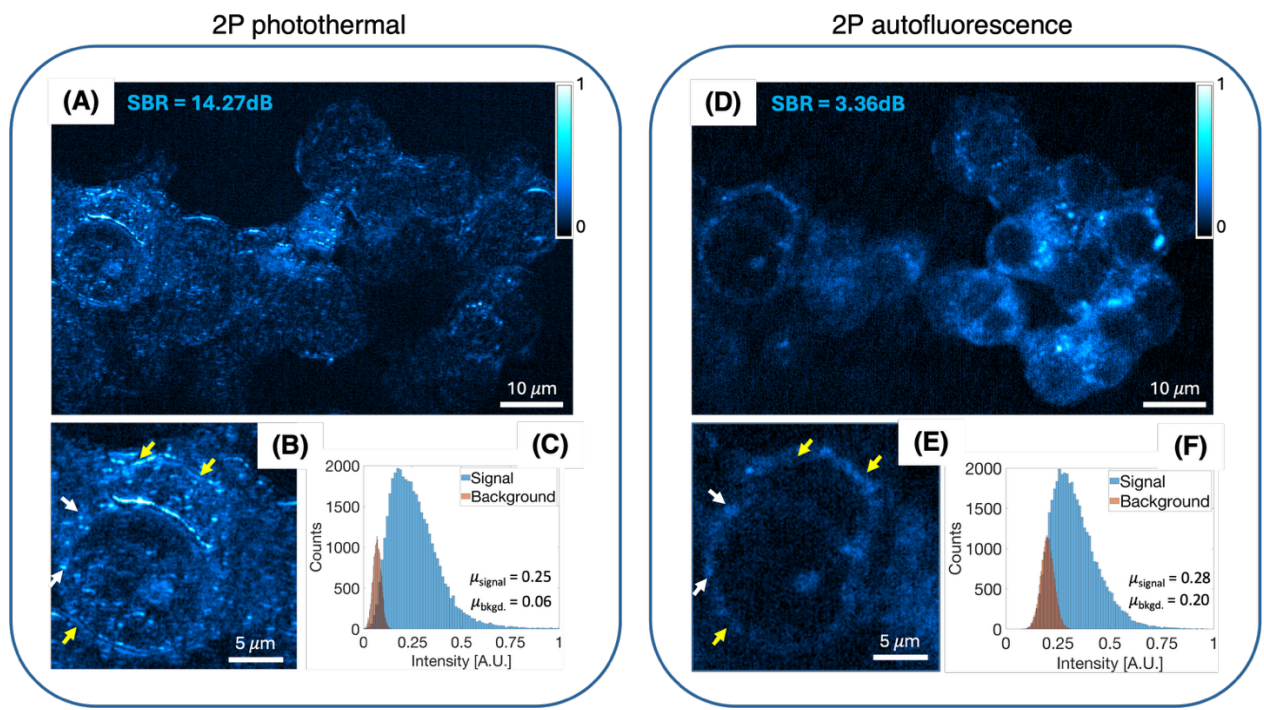

**Figure 2. Comparison between 2PPT and 2PAF for imaging mitochondria NADH in cancerous cells.** (A), (D) Monolayer of fresh, energy producing SK-OV-3 cells obtained with 2PPT and 2PAF, respectively. (B), (E) A single cell image showing tubular mitochondria (yellow arrows) and blob/oval mitochondria (white arrows) obtained with 2PPT and 2PAF, respectively. (C), (F) Intensity histogram of images obtained with 2PPT and 2PAF, respectively, that shows signal (blue bins) and background (orange bins) along with the mean values denoted by $\mu_{signal}$ and $\mu_{bkgd.}$.

Unlike labeling optical microscopy approaches that utilize chemical probes to target the biomolecule of interest, 2PPT rely on local heat generation to provide the biomolecular contrast and thus lacks molecular specificity. Therefore, to further validate the contrast origin in 2PPT, we studied the spectral response of the observed subcellular features in the tested cells and compared it with pure sample standards. Other than metabolic coenzymes in living cells, heme proteins are considered major chromophores for two-photon absorption in the NIR region that can potentially contribute to the local heat generation (37,48). Therefore, we studied the 2PPT spectral response of cytochrome C in addition to the metabolic coenzymes of interest NADH and FAD. In this experiment, the pump beam, which dictates the two-photon absorbance of the specific chromophore, was swept across the first NIR 'biological window' relevant for subcellular features (49). **Figure 3(A)** plots the 2PPT signal of NADH solution (0.25 mM) as a function of pump beam wavelength. As expected, a strong 2PPT signal is observed around 720 nm that originates from the peak in two-photon absorption of the chromophore (43,44) and decrease thereafter. Notably, an even stronger signal is observed below 700 nm with pronounced dip precisely at 700 nm. We attribute this behavior to the involvement of two-photon excitation of the NADH biomolecules to the second excited state, which exhibit a very prominent one-photon (1P) absorption at the UV-C region (44). This higher, almost double, signal presumably originates from combination of the UV-A tail (1P excitation at ~355 nm) which overlaps with the stronger UV-C contribution (1P excitation at ~260 nm) associated with transition to a higher electronic band ($S_2$) (50). **Figure 3(B)** shows the spectral response for 0.25 mM FAD solution, where a relatively homogenous 2PPT signal is observed over extended bandwidth 770-820 nm due to a solid 2P absorption in this range. Given its prominent 1P

Page **7** of **24**

absorption at UV-C wavelengths, FAD spectrum also exhibits a strong signal below 700 nm, albeit with a less pronounced dip due to a weak absorption around 720 nm (by contrast to NADH). The 2PPT spectral response of cytochrome C (0.25 mM solution), shown in **Figure 3(C)** and obtained for similar bandwidth as the NADH and FAD solutions, shows significantly different spectral features with a strong signal at 750-760 nm that steadily decays immediately afterwards as it reaches 900 nm. To facilitate quantitative comparison between the spectra shown in **Figure 3**, we normalized each spectrum with respect to the water absorption at 960 nm since each sample has the same solute concentration (51). The observed 2PPT spectroscopy of cytochrome C agrees with the strong single-photon absorption it has around 400 nm (52), by contrast to the weaker absorption by both the metabolic cofactors NADH and FAD in this region that is expressed by the low 2PPT signal around 750 nm. To determine the identity of main contributor to heat generation when 2PPT is employed in subcellular structures, we obtained spectral measurement of viable SK-OV-3 cells with different 2PPT pump wavelengths within the relevant window (see Movie S1). **Figure 3(D)** shows the spectral response of 20 randomly selected mitochondria visible in the subcellular compartments. By inspecting the 2PPT spectra, one can appreciate the common spectroscopic features between the metabolic cofactors and the mitochondria. Namely, the declined signal (dip) around 750 nm and the relatively constant signal that starts at 770 nm and extends to 820 nm. These features are missing in the heme-protein spectrum. Notably, the strong signal below 700 nm that is observed in the solution samples is not observed in the cells. This is explained by decreased absorption from the adenine ring in the UV-C band, which occurs when the biomolecule is in bound-state compared to its free-state in aquatic samples (53). The similarity between these randomly selected organelles and the FAD spectral shape further suggests that under our experimental conditions, most of the signal associated with 2PPT originates from the metabolic coenzymes rather than the heme protein. To provide quantitative evaluation of this similarity, we compared the four spectra shown in Figure 3 with phasor analysis, that compares their spectral shape regardless of intensity by projecting it to the complex phasor space (54). We only considered the 2PPT-relevant band of the spectrum (UV-C and water-absorption excluded), and we found that the closest spectrum to that of mitochondria was FAD followed by NADH while that of cytochrome C was the least similar (see Figure S4). Nevertheless, the subcellular contrast in photothermal imaging modalities is highly dependent on the experimental conditions (cell line, culture and preparation protocol, observation time, etc.) and is still an open debatable question. For example, it was previously shown that mitochondria are the major source of heat generation under one-photon excitation in a different type of cells due to an ensemble of proteins (48).

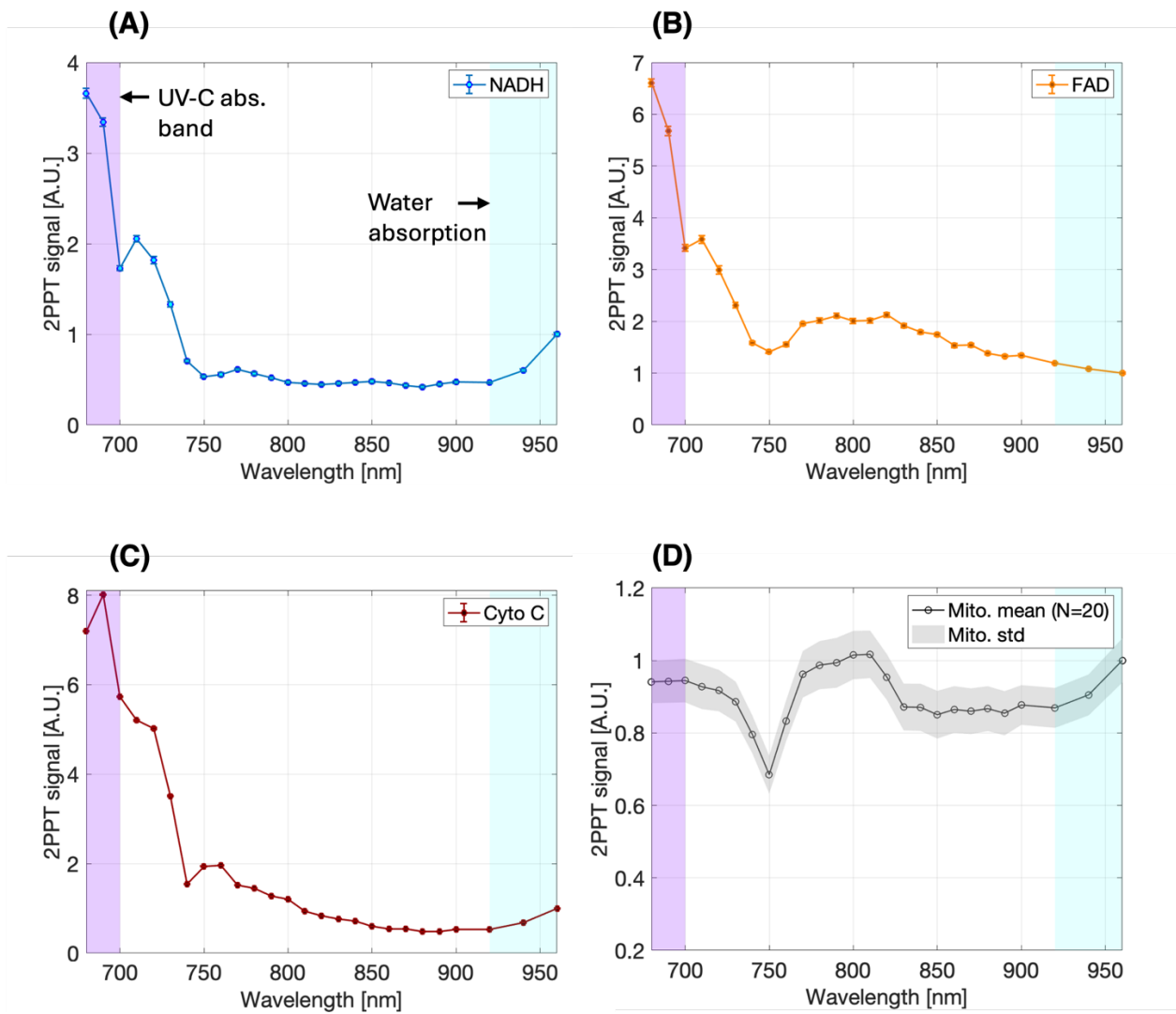

**Figure 3. Validation of contrast source from mitochondria.** 2PPT spectroscopy of (A) NADH, (B) FAD, (C) cytochrome C, and (D) mitochondria from live SK-OV-3 in the range 680-960 nm. 2PPT spectroscopy of mitochondria shows similar spectroscopic features to the FAD coenzyme (dip around 750 nm, consistent signal between 770-820 nm). Bands associated with UV-C absorption and water absorption are highlighted by light purple and blue, respectively.

## 2PPT imaging of metabolic perturbations in ovarian cancer cells

Metabolic perturbations are hallmark of cancer, reflecting the profound reprogramming of cellular energy pathways that supports uncontrolled proliferation, survival under stress, and resistance to therapy (55). Investigating these perturbations under controlled conditions, such as nutrient deprivation and inhibition of reactive oxygen species (ROS), is essential for elucidating the adaptive mechanisms that enable cancer cells to survive within hostile microenvironments. The following experiments show 2PPT capability to image dynamic metabolic modifications in cells that stem from biosynthesis of NADH and FAD during cancer cells' growth and proliferation.

We first studied the effects of ROS inhibition on the cells capacity to proliferate and produce energy. For this end, SK-OV-3 cells were cultured and grown as a single layer on a glass-bottom dish as previously described. The treatment group consists of cells that were treated with N-acetylcysteine (NAC), which inhibits the production of ROS that are vital for energy production. **Figure 4(A)** and **Figure 4(B)** illustrate 2PPT micrographs obtained with 720 nm pump beam (targets NADH) of control and NAC-treated cells, respectively. Box plot shown in **Figure 4(C)** confirms the significantly reduced NADH levels in the NAC-treated cells compared to untreated control based on the measured SBR. Similarly, **Figures 4(D)-4(F)** show consistent trends derived from FAD expression in cells, obtained by tuning the pump beam to 780 nm. We observed significantly lower amounts of both coenzymes in the NAC-treated cells compared to control cells, with 33% and 37% reduction in NADH and FAD expression (median SBR value), respectively. Notably, the NAC-treated cells exhibit no mitochondria in the outer nuclear membrane compared to their existence in the untreated cells. Moreover, while abundance of rod-shaped mitochondria is observed in the untreated cells [**Figure 4(A)** and **Figure 4(D),** yellow arrows], the ROS-inhibited cells show minimal (to not at all) rod-shaped microstructures in both channels and more oval-shaped micrometer scale mitochondria within the cytoplasmic region [**Figure 4(B)** and **Figure 4(E),** white arrows]. These observations of lower energy production accompanied with mitochondria shape change might highlight an ongoing regularization process of the cells to compensate its dynamic energy needs (56,57).

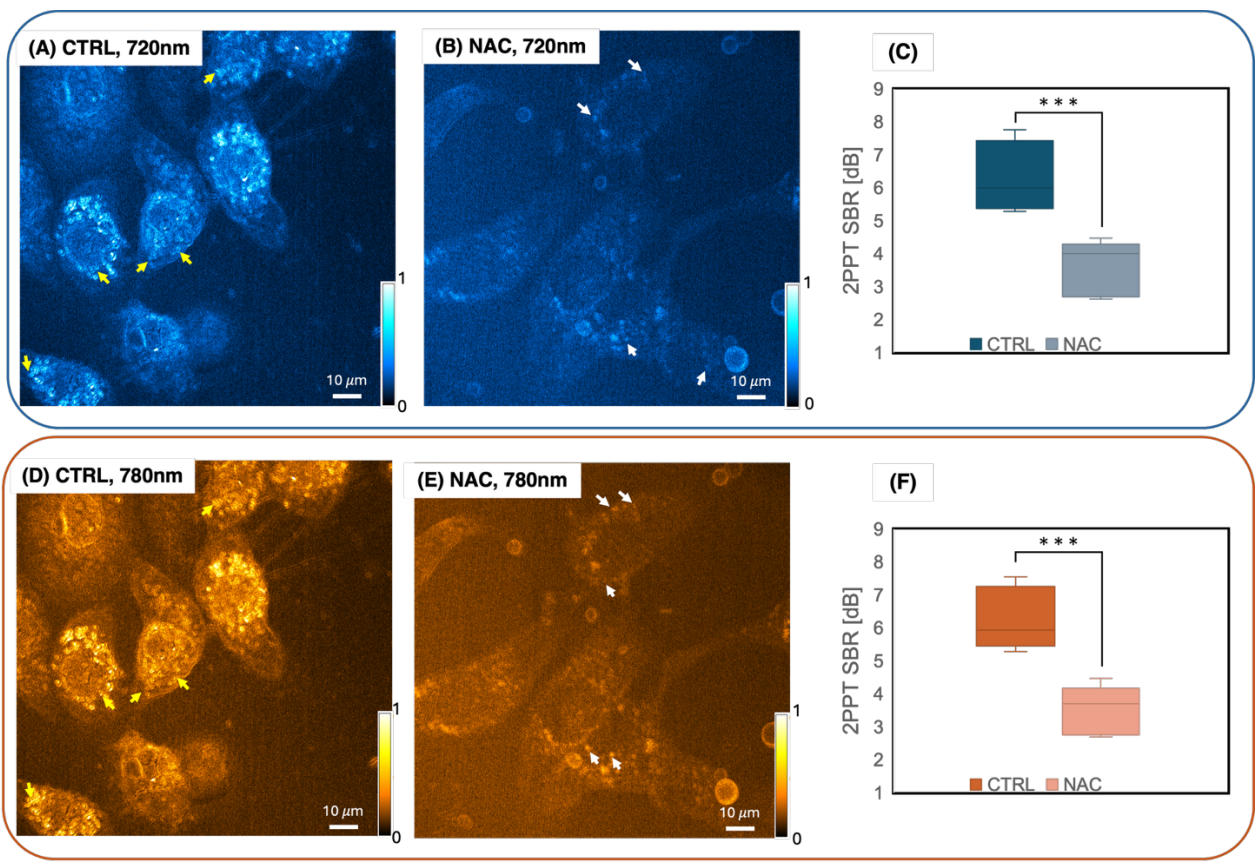

**Figure 4. 2PPT microscopy captures metabolic response of SK-OV-3 cells to ROS inhibition.** 2PPT micrographs obtained (A) without and (B) with N-acetylcysteine (NAC) treatment, captured with pump of 720 nm targeting NADH. Significant reduction in NADH coenzyme production is observed in (C). (D), (E) shows equivalent images to (A), (B) for the pump at 780 nm targeting FAD. Significant reduction in FAD coenzyme is observed in (F). Yellow and white arrows in micrographs mark regions of rod-shaped and oval-shaped mitochondria, respectively.

We further expanded our investigation of the metabolic reprogramming of cells under intentional disruption of energetic balance using 2PPT and incubated the cells to grow under starvation conditions for 24 and 48 hours (see Materials and Methods for more details). Compared with the equivalent control groups, that were supplemented with sufficient nutrients during incubation, the starved cells activate protective responses, including autophagy to recycle internal components for energy and cell cycle arrest to reduce energy demands. **Figure 5** summarizes the starvation experiment results for 24h and 48h as captured by both NADH and FAD channels in 2PPT. Data collected for NADH and FAD are grouped into separate panels **Figure 5(A)** and **Figure 5(B)**, respectively. Each panel contains a representative single cell image from each condition in the experiment as labeled (24h control and starved, 48h control and starved). Few selected tubular mitochondria and oval mitochondria are marked in left-most single cell image by yellow and white arrows, respectively. Bar plots in **Figure 5(A)** and **Figure 5(B)**, below the single cells frames, summarize our findings following the dual timepoints starvation experiment. To gain further insights into how these cells change their energy consumption and production, we divided the data obtained from N=6 cells per condition (full field of view data in Figure S5) to signal that comes from tubular (commonly healthy) mitochondria and oval shaped (commonly unhealthy) mitochondria (45,46). To perform this task, we employed image segmentation followed by shape and size analysis. Further details on how both types of mitochondria were identified and categorized can be found in Materials and Methods under subsection 'Data analysis'. Data collected from tubular mitochondria within the cells show statistically significant increase in NADH (median SBR) levels following their starvation for 24h. Further starvation of these cells for 48h reveals a decline in NADH levels measured from tubular mitochondria. Although this difference did not reach statistical significance in our reported data set (p=0.07, **Figure 5(A)**), the directionality of the change is consistent with a transient metabolic adaptation at 24h, potentially reflecting increased energetic allocation to maintain cellular processes, followed by a shift toward a more energy-conserving state at 48h. On the other hand, the NADH production from oval mitochondria follows a different trend with significant signal decay (median SBR) between the untreated and treated cells for both 24h and 48h starved cells. This suggest that the presumably unhealthy mitochondria are minorly associated with energy production that further decline when cells are under stress (starvation) conditions. Examination of the FAD related signal from the same location shows non-significant change in the expression of the FAD levels (median SBR) from the tubular mitochondria. Meanwhile, the observed trends from oval mitochondria for FAD expression follow similar trends of the NADH production, which might indicate that these dormant compartments are inactive during the cellular crisis and potentially have lost their functionality.

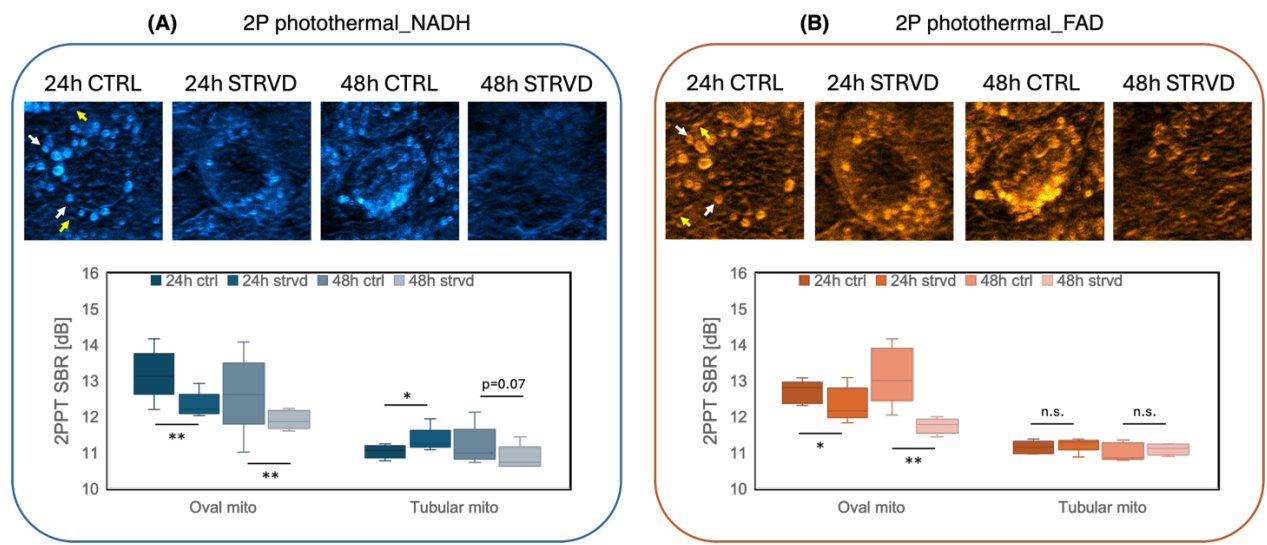

**Figure 5. 2PPT microscopy visualizes different mitochondria morphologies in starved SK-OV-3 cells via NADH and FAD sensing.** (A) Representative single cell images (full field images in Figure S5) after 24h and 48h starvation along with their controls obtained from NADH channel with pump 720 nm. Box plot below shows the trends observed for oval mitochondria (white arrows in 24h CTRL) and tubular mitochondria (yellow arrows in 24h CTRL). (B) Shows equivalent data to (A) obtained from FAD channel with pump 780 nm.

**2PPT imaging of ovarian cancer spheroids response to chemotherapy**

Tumor spheroids derived from ovarian cancer cells provide a physiologically relevant platform to investigate metabolic alterations under conditions that closely mimic the in vivo tumor microenvironment. Within these three-dimensional constructs, gradients of oxygen, nutrients, and pH establish distinct metabolic niches that drive heterogeneity in cellular energy production and utilization (58–60). Characterizing these spatial metabolic alterations is therefore crucial for understanding how ovarian cancer cells adapt their mitochondrial function and redox balance to sustain proliferation, survive therapeutic stress, and develop resistance. NADH and FAD serve as intrinsic metabolic reporters whose optical signatures can reveal shifts in the balance between glycolytic and oxidative pathways across the spheroid architecture (22). Therefore, to investigate how cancer progression might be suppressed under chemotherapy treatment and assess the underlying mechanisms of tumor adaptations toward resistance, we study cancer spheroids under cisplatin drug treatment with the 2PPT microscope. Cultured SK-OV-3 ovarian cancer cells were supplemented in supporting tubes to facilitate their growth into spheroids of ~500 μm in diameter (see Materials and Methods for further details). Spheroids were then passaged and suspended in aquatic medium for imaging. Of note, 1P absorption of water from the 3$^{rd}$ overtone in the ~720 nm band is one order of magnitude weaker than its 2$^{nd}$ overtone in the ~970 nm band (51) and does not interfere with our 2PPT metabolic readouts, as demonstrated by the spectroscopic measurements of aquatic solutions and cells (Figure 4 and Movie S1). Maximum amplitude projections (MAP) of 2PPT micrographs along 100 μm deep volume of untreated (labeled CTRL) and cisplatin-treated (labeled CIS) spheroids from both NADH and FAD channels [**Figure 6(A)-(B)**, **Figure 6(D)-(E)**], together with their respective 2PPT SBR statistics are summarized in box plots [**Figure 6(C), 6(F)**]. From the depth projected images of both NADH- and FAD-targeted photothermal maps, one can observe the nucleus membrane and cytoplasmic regions of individual cells constructing the tumor-mimicking spheroid. Inspection of the NADH map shows a sharp decline in intensity for cisplatin-treated spheroid, while the equivalent FAD-associated map shows an opposite, albeit smaller, trend, with increased signal for the treated spheroid. These observations are further confirmed by comparing 2PPT signal from the cytoplasmic regions of 5 cells per each spheroid. Statistically significant signal reduction of NADH is observed after the chemotherapy treatment [**Figure 6(C)**], while stronger signal is observed for the FAD [**Figure 6(F)**]. Note that 2PPT SBR is reported as the signal levels in box plots [**Figure 6(C)** and **Figure 6(F)**] to facilitate comparison with other experimental data shown in this work (e.g., starvation test). The trends observed herein highlight the adaptability of cancerous cells to respond an ongoing adversarial action from the hosting tumor. The different balance between the two metabolic coenzymes in untreated spheroids versus the treated ones, may be a marker of ongoing metabolic adaptation within the tumor to facilitate its growth.

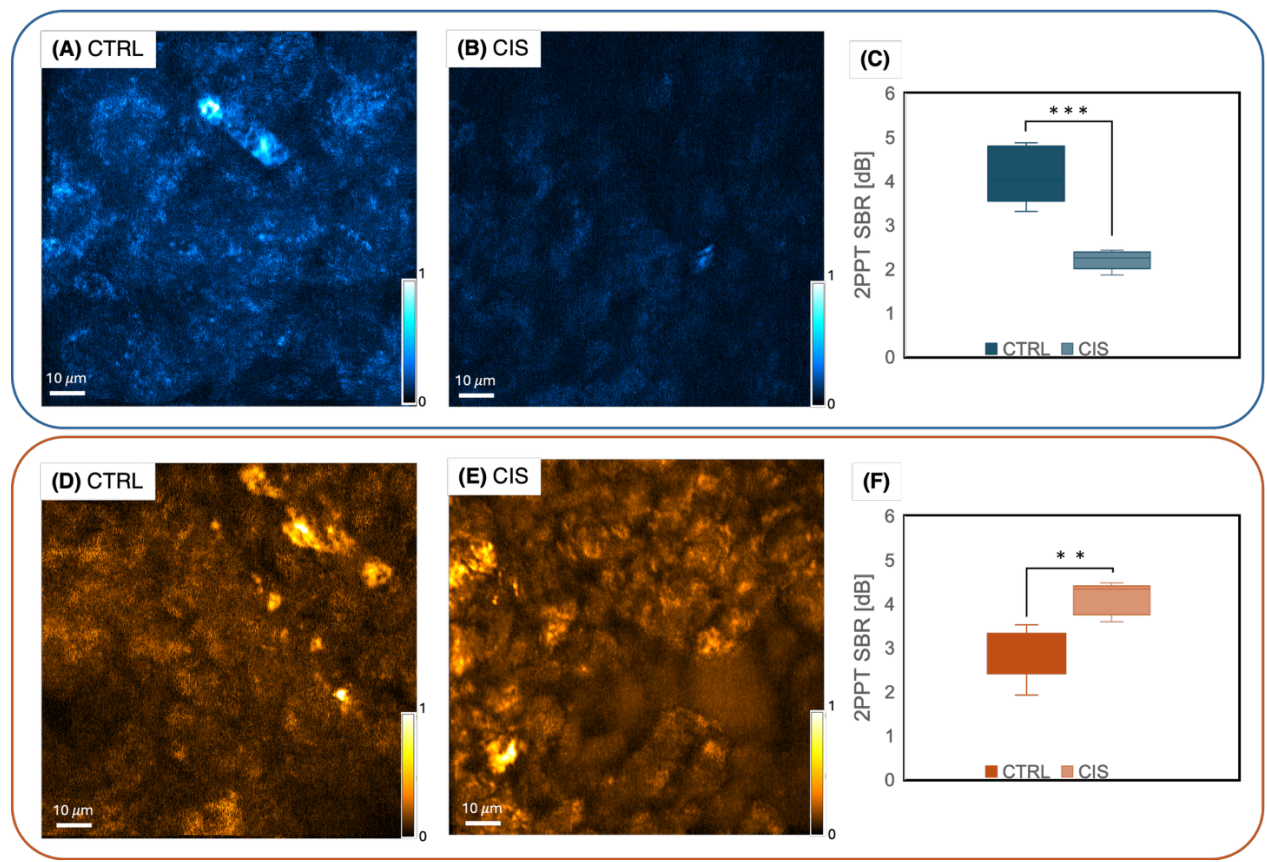

**Figure 6. Metabolic heterogeneity in cancerous spheroids after chemotherapy treatment captured by 2PPT microscopy.** Maximum amplitude projections (MAPs) of SK-OV-3 spheroids obtained with 2PPT for (A) untreated and (B) cisplatin-treated conditions captured with pump of 720 nm targeting NADH. Significant reduction in coenzymes production is observed in (C). (D), (E) shows equivalent images to (A), (B) for the pump at 780 nm targeting FAD. Significant increase in FAD coenzyme is observed in (F). CTRL: untreated, CIS: cisplatin-treated.

## Discussion

It is well established that the pathogenesis and progression of carcinomas are linked to alterations in the metabolic activity of epithelial cells (61). The extensive tumor needs to supplement rapid growth, survival and adaptation are met by significant metabolic reprogramming of the cells, of which the Warburg effect, glutaminolysis and lipogenesis are well-known hallmarks (62,63). These metabolic disorders have long been studied as predictive agents of oncogenic onset since they are associated with concomitant modification of the native cells environment at the (sub)microscale (64–67). In fact, the metabolic conditions of cells and tumors strictly influence the amount, distribution and biochemical state of the endogenous chromophores within the biological system (11,12,43). Notably, NADH and FAD, which are almost the sole source for native fluorescence arising from cell cytoplasm, were shown to carry important information about carcinogenesis and disease progression. Importantly, the metastatic potential and disease stage are closely related to the microstructural and biochemical state of the precancerous cells as captured during the NADH and FAD consumption and production rendering these coenzymes an important cancer biomarker (3,68). These results further fuel the ongoing efforts to target cellular metabolism to improve cancer therapeutics. The links between dysregulated cellular metabolism and cancer drug resistance hold great promise to enhance the efficacy of common therapeutic agents (69) and design new combination therapies (4,70,71). Many of the mentioned findings were enabled by the perseverant fusion of interfacing areas of research commonly addressed to as biomedical imaging.

A key aspect of a successful microscopic technique for life science is the ability to extract biomedically relevant contrast in a label-free and non-destructive manner. Inspired by this notion, in this study we demonstrate the potential of two-photon photothermal (2PPT) microscopy as a sensitive and specific label-free imaging modality to probe metabolic-relevant biomolecules within living systems. By sensing the photothermal response of metabolic cofactors upon their two-photon excitation, our approach surpasses conventional label-free fluorescence-based imaging in its ability to spatially map cellular metabolic activity with via mitochondria imaging. Traditional autofluorescence techniques that rely on photons emission from NADH and FAD often suffer from limited sensitivity due to background scattering, spectral overlap, and low quantum yields constraining their accuracy in detecting subtle metabolic alterations. The experimental demonstrations presented herein establish 2PPT capability of enhanced SNR and precise localization of mitochondrial metabolism, providing a new analytical window into dynamic redox states within living cells and 3D culture systems.

We first validated 2PPT sensitivity to endogenous metabolic biomolecules by demonstrating its sub-micromolar LOD, which is 10- and 20-times more sensitive than the well-established, widely used autofluorescence approach for FAD and NADH, respectively (25,27). This improvement in sensitivity has allowed us to visualize the heterogeneity of mitochondria compartments within cancer cells highlighting the relationship between its structural variability and the cells redox state in a label free manner (56,57). We identify spectroscopic measurement in 2PPT microscopy as a distinct feature that can provide chemical insights about the probed biological environment, decoupled from fluorescence emission inconsistencies. This was utilized in our studies to verify that 2PPT signal from organelles mainly originates from metabolic co-factors (**Figure 2**). Importantly, the ability of our approach to detect perturbations in metabolism highlights its robustness and specificity in capturing biochemical response. By intentionally disrupting the cellular metabolic activity (e.g., ROS inhibition and nutrient starvation), we demonstrate that 2PPT imaging accurately reflects intracellular alterations through distinct variations in the photothermal signatures of NADH and FAD. These findings establish a direct link between the optical contrast generated by our system and the functional metabolic response of the cell, confirming that our apparatus can serve as a quantitative, label-free probe of bioenergetic adaptation. This sensitivity to transient metabolic shifts positions 2PPT microscopy as a valuable complement to established biochemical assays, offering real-time, subcellular-level insight into metabolic regulation.

Imaging biomolecules from deep within three-dimensional constructs that scramble light propagation is a long-standing effort in translational imaging (72,73). While we have not rigorously studied how 2PPT signal scales compared to other fluorescence-based techniques in extraction of bulk-tissue interactions, we do recognize the potential it has to surpass current methodologies in achieving greater effective penetration. Spatial resolution in 2PPT is determined by the spatial overlap between the pump and probe beams. Meanwhile, the nonlinear nature of the 2P excitation provides inherent optical sectioning and therefore 3D imaging capability. Since effective photothermal process can be induced with non-diffraction-limited excitation beam, 2PPT can in principle be implemented with a probe beam that minimizes the difference in aberrations compared with the pump beam within the scattering medium and register meaningful readouts deeper than fluorescently emitted photons. Overall, 2PPT, as a highly sensitive transient-based approach, can potentially register metabolically relevant information even in case of severe distortion to its wavefront. Much like photoacoustic microscopy sensors (74), 2PPT has the capacity to sense relevant data as long as sufficient excitation is provided – and this is decoupled from the probe beam characteristics. Specifically, to leverage the three-dimensional imaging capability of our approach, we extended our analysis to multicellular ovarian cancer spheroids to evaluate metabolic heterogeneity and treatment response. Within these physiologically relevant constructs, we observed a downregulation of NADH expression concurrent with an upregulation of FAD following chemotherapy exposure, reflecting a transition toward an oxidative metabolic phenotype. These results align with previous observations of redox remodeling in drug-treated cancer systems (75,76). The capacity to resolve such metabolic gradients across spheroidal volumes demonstrates the advantage of 2PPT over conventional fluorescence microscopy, which is often limited by scattering and photobleaching in dense biological samples.

Our current contribution follows a recent path of research that investigates approaches to develop novel, label-free optical techniques for probing cellular metabolism, complementing and extending previously established modalities. Historically, autofluorescence imaging, particularly two-photon excitation of NADH and FAD combined with fluorescence-lifetime measurements, has served as a foundational approach to map redox state and metabolic dynamics in living cells and tissues (3,7,11,12,14,43). Photoacoustic imaging represents another powerful label-free route that achieves deep penetration and maps chromophores in thick tissues (77). More recently, multiphoton photoacoustic microscopy has been used to detect endogenous NADH deep in brain tissue (78). Additionally, a novel polarization ratiometric microscopy enables rapid, high-resolution tracking of NADH anisotropy as a functional readout of metabolic state without requiring lifetime measurements (79). These advances underscore a broader shift: combining absorption, thermal, acoustic, and polarization contrasts to report metabolism in more diverse, deeper, and more physiologically relevant contexts. Our 2PPT approach offers a distinct and complementary capability. By sensing photothermal signatures associated with NADH and FAD, we combine strong molecular sensitivity with multiphoton excitation's spatial resolution and 3D imaging compatibility. This positions 2PPT as a valuable, label-free addition to emerging metabolic imaging platforms – bridging the gap between fluorescence-based lifetime methods, ultrafast absorption, and deep-tissue photoacoustic approaches.

While our study provides initial validation of 2PPT imaging for metabolic sensing, several aspects warrant further investigation, including optimization of detection parameters, calibration of photothermal signals with absolute metabolite concentrations, and extension to in vivo models. For instance, the direction and magnitude of NADH and FAD changes after therapy are sensitive to drug mechanism, dose and imaging timepoint – some reports show transient NADH increases (or decreases) depending on whether cells undergo early OXPHOS activation, oxidative stress, or apoptosis (75). Therefore, cross-study comparisons require careful matching of experimental conditions. Future studies should aim to integrate 2PPT with complementary imaging modalities and develop analytical models to correlate photothermal signatures with metabolic flux. Such efforts will be crucial for translating this methodology toward clinical and diagnostic applications in metabolic and oncologic research.

**Materials and Methods**

Simulation of energy deposition and heat propagation to optimize 2PPT throughput

To evaluate the feasibility and enhance the registration of meaningful metabolic signal after optical heat deposition by the pump beam, we carried out a simulation to mimic the photothermally generated transients. Our simulation goals were to 1) verify thermal lensing, both spatially and temporally, within the impacted volume and to 2) estimate needed relaxation time for sufficient heat buildup. These further helped us to optimize the needed pump and probe beams characteristics. Our chosen tested medium was 0.5 mM NADH solution. After calculating the amount of initial temperature rise deposited to the medium via $Q/\rho C_p$, with Q being the total amount of absorbed energy, $\rho$ the molecule's density and $C_p$ the material's heat capacity, our simulation started by solving the second order partial differential equation known as the heat equation:

$$(1) \frac{\partial}{\partial t} T(\boldsymbol{r}, t) = \alpha \nabla^2 T(\boldsymbol{r}, t) + \frac{Q}{\rho C_P}$$

where $\alpha$ is the thermal diffusivity, and T is temperature. Eq. (1) was solved for two cases during a single excitation-relaxation process: with a nonzero deposited energy term from the pump beam within the impacted volume, and with a zero deposited energy outside that volume. The simulation was done on a two-dimensional grid (X-Y) mimicking transverse heat dissipation across the sample at the focal plane and used Dirichlet boundary conditions to account for the fixed surrounding temperature. For simplicity, we considered a flat-top pump beam of ~0.5 μm diameter, which corresponds to the Airy disc of the used wavelength and optics, and a uniformly absorbed energy Q. Figure S1 describes the temperature field within the solution during the excitation of a single pulse and multiple pulses, from which an optimal number of photothermal events was calculated to set the experimental pump beam modulation (see subsequent section and Supplementary Material). In addition to time-resolved heat generation, Figure S1 shows heat map in different time points where a radially decaying temperature field is observed. Our simulation emphasizes that, in our current 2PPT implementation, heat accumulation from multiple pulses is required to obtain a sufficient signal that is detectable by intensity measurements of the probe beam divergence (or convergence) through the thermal lens.

Two-photon photothermal (2PPT) microscope

Our 2PPT microscope is based on acousto-optic modulated (AOM, 1205C-843, Isomet Corporation, Manassas, VA) pump beam (tunable, 680 to 960 nm) and an unmodulated probe beam (522.5 nm frequency doubled via a Lithium Triborate crystal (LBO-604H, EKSMA Optics, Vilnius, Lithuania), originally from 1045 nm seed laser), both of which originate from an 80 MHz fs solid state ultrafast laser oscillator (InSight X3, Spectra-Physics, Andover, MA). To maximize the stability of the generated probe beam, the nonlinear crystal was illuminated with a focused beam and high peak power that was suppressed before interacting with the sample. Both beams collinearly directed to the back focal plane of a 60x 1.2 NA water immersion microscope objective (Olympus, Japan) after steered by a pair of galvos (GVS002 ,Thorlabs, Newton, NJ) to raster scan the sample of interest. On the sample, the probe beam size and the probed volume thickness were approximately 350 nm and 10 μm, respectively. After interacting with the sample (dwell time=20 μsec), probe beam signal was collected via a condenser lens equipped with an iris to dynamically sense the photothermal lensing (80) and directed to a photodetector (DET100A2, Thorlabs, Newton, NJ) via a 525 nm bandpass optical filter (CT525/30, Chroma Technology Corp., Rockingham, VT). Average optical powers were set to 12 mW (pump beam) and 7 mW (probe beam), measured at the sample plane (no phototoxicity was observed). To enhance signal, our photodetector was equipped with 50-ohm resistor, a 22-kHz high-pass radio frequency filter, and a 46-dB low-noise amplifier (SA230-F5; Wayne). The photodetector (10 MHz bandwidth) output was sent to a lock-in amplifier (MFLI, Zurich Instruments USA, Inc., Waltham, MA) that samples the signal at 60 MHz to satisfy Nyquist criterion and the 2PPT signal was recorded with a data acquisition card (NI-DAQ, PCI-6363, National Instruments, Austin, TX). The 2PPT system was further synchronized by the NI-DAQ card, which controlled the galvos scanning unit and the function generator that feed the AOM. To allow optimal energy deposition and heat generation, the function generator was set to rectangular wave modulation at 125 KHz and 5% duty cycle (no pulse shaping is involved). The amplified signal from the photodetector was demodulated by the pump beam reference and sent to a PC station to render and record the image.

Preparation of NADH/FAD samples for characterization and LOD test

To characterize the 2PPT microscope, pure NADH and FAD samples were prepared in 1 mM stock solution from purchased high purity cofactors powders (β-Nicotinamide adenine dinucleotide, N8129; Flavin adenine dinucleotide, F6625, MilliporeSigma, Burlington, MA) by mixing it with phosphate buffered saline (PBS pH 7.4 1x, Gibco, Thermo Fisher Scientific, Waltham, MA) to achieve desired concentration. For LOD tests, titration was made by halving concentration of stock solution. To maintain a high accuracy of concentrations, a one-time 100 mM master solution was first made and then serially diluted to obtain the reported concentrations in the micro-molar range. For each concentration in the LOD curve, a 1.3 μL of solution was sandwiched between two 1.5# glass coverslips and mounted at the objective's back focal plane. The sample thickness was controlled by a ~250 μm spacer used during the droplet encapsulation. Imaging medium was water and collection medium was oil. Data was obtained by imaging the droplet center with the 2PPT and 2PAF modalities, and signal was extracted from a 5 $\mu$m diameter centroid region (data not shown). Signal was then plotted against the concentration, and the best linear fit was calculated using MATALB R2023a curve fitting tool (MathWorks, Inc., Natick, MA). For each condition, the residuals standard deviation ($\sigma$) and the fit slope ($m$) were extracted, and the LOD was determined according to LOD=3.3·$\sigma$/$m$, which is a standard approach for linear regression to calculate the value that represents when the signal is significantly different from the noise (42).

Cell culture and spheroids preparation

SK-OV-3 cells (American Type Culture Collection) for both 2D and 3D culture were cultured in the DMGM Basal Medium (Cell applications, MCBD 105 Medium and Corning, Medium 199 1X combined 1:1) supplemented with 10% fetal bovine serum (FBS) and penicillin/streptomycin (P/S; 100 U/ml) and maintained in an incubator with 5% $CO_2$ and 80-90% humidity at 37°C. For 2D cell culture samples, 24 hours before imaging, 75,000-100,000 cells were seeded in 35 mm glass bottom dishes for control and treated groups in the complete DMGM medium. For treatment groups with N-Acetyl-L-cysteine (NAC, Millipore Sigma Aldrich, A91654), the cells were incubated with 5 mM NAC for 1.5 hours before imaging. For fluorescence imaging, the cells were incubated with Mito tracker (MitoTracker Green FM, M7514 ,Thermo Fisher Scientific, Waltham, MA,) at a concentration of 100 nM for 15 min before imaging. For starvation experiments, the cells were seeded in the complete DMGM medium for 24 hours followed by 24- or 48-hours incubation in DMGM supplemented with 1% (and 0.5% FBS) for different starvation levels. For the live cell imaging, the glass bottom dishes with cells were washed once gently with 1X PBS, followed by immersion in 1X PBS. For the 3D spheroids imaging, 8,000 SK-OV-3 cells per well were seeded in ultra-low attachment round-bottom 96-well plates and incubated for 48 hours till the spheroids formed in the complete DMGM medium. For treatment groups with cisplatin (232120, MilliporeSigma, Burlington, MA), the spheroids were incubated with a final concentration of 10 $\mu$M for 4 hours. For the live spheroids imaging, the spheroids were washed twice gently with 1X PBS and then transferred to 35 mm glass bottom dish with 1X PBS.

Two-photon autofluorescence (2PAF) setup

To facilitate comparison with 2PPT and ensure localization of subcellular features between the two methods, 2PAF measurements were obtained using the 2PPT setup illustrated in **Figure 1(A)** with the necessary modifications, as follows. The tunable port from our ultrafast laser was used as the light source for 2P excitation and was directed toward the sample using the same optics (AOM modulation off). To allow collection of fluorescence signal in reflection mode, a dichroic mirror (DM, #69-899, Edmund Optics, Barrington, NJ) was introduced between the microscope objective and the galvos scanners. Emission was collected via the illumination objective with 10 μsec pixel dwell time and was directed towards a photomultiplier tube (PMT, H16722-40, Hamamatsu, Japan) after passing through a focusing lens and a chromatic filter to pick chromophore-specific signal 460/10 for NADH and 520/10 for FAD (FBH460-10; FBH520-10, Thorlabs, Newton, NJ). Current signal from the PMT was converted to measurable voltage by an in-house built transimpedance amplifier equipped with 1 MHz low-pass filter and sent to the digitizer with no further operation. Two-photon excitation was verified by imaging 2.9 $\mu$m green-fluorescent beads (Fluoro-Max, G0300, Thermo Scientific) and confirming non-linear scaling of the signal with the excitation power.

Data analysis

Comparison of imaging performance between 2PPT and 2PAF modalities was done by selecting 25 patches of 4 $\mu$m x 4 $\mu$m from cells-containing area and cells-free area in acquired micrographs (**Figure 2**) to be accounted as the signal and background data, respectively. After manual classification of the micrograph regions to cells (signal) and cells-free (noise), the patches from each class were randomly selected to avoid potential bias. We then analyzed the signal and noise histograms (**Figure 2**), from which SBR was calculated in dB based on the means approach as $10 \cdot \log(\text{mean}_{signal}/\text{mean}_{background})$. Spectral response of mitochondria features to verify source of contrast in 2PPT [**Figure 3(D)**] were collected from stack micrograph of cells obtained for pump beam swept in the range 680-960 nm (see video in Supplementary Materials). 20 different regions of 5 $\mu$m x 5 $\mu$m containing different shapes of mitochondria were collected from 6 SK-OV-3 cells and the mean and standard deviation of 2PPT signal was used to plot the mitochondria spectral response curve. Similarly, statistical data of 2PPT from subcellular mitochondria that was used in metabolic perturbation and chemotherapy assessment experimentation [**Figures 4-** 6] was obtained. Minimum of 7 mitochondria-containing regions of 5 $\mu$m x 5 $\mu$m were registered per single cell and data from total of 5-6 cells were averaged per experimental condition (including control). To differentiate between rod-shaped and oval-shaped mitochondria, we first discriminated the subcellular organelles from their background by applying a Gaussian smoothing filter and followed by an automated bi-modal histogram-based image thresholding algorithm to maximize the separability of the resultant classes of noise and signal pixels (81). We then created a binary mask where mitochondria were defined as pixels above the automatically selected intensity threshold value. The different mask features were characterized based on their shape, with a custom-built algorithm to identify elongated and oval structures in the image data. Finally, the classified features mask was merged in the signal extraction pipeline described above to report the signal values of the two mitochondrion types (**Figure 5**). All these steps were specifically developed for this study and were implemented in MATLAB (R2023a, MathWorks, Inc., Natick, MA). To visualize spheroids data obtained from 100 $\mu$m thickness volume of the interrogated constructs we used the maximum intensity projection method. A custom-built MATLAB code to align the different slices separated by ~20 $\mu$m was used to account for lateral distortion and sampling errors between the registered planes, followed by projecting the maximum value per lateral location within the volume onto the two-dimensional micrographs to obtain the MAP (**Figure 6**).

Statistical analysis

The statistical graphs were shown as means ± stdev unless specified otherwise. The statistical analysis was done using either unpaired t-test between treated and untreated groups. Statistical significance was denoted as * for $p<0.05$, ** for $p<0.005$, *** for $p<0.0005$ and n.s. for $p>0.05$ indicating a non-significant statistical difference.

**Acknowledgments**

**Funding**

This work was supported by NIH grant R35 GM136223 (JXC).

**Author contributions:**
Conceptualization: NH and JXC
Methodology: NH, CVPD, JA, DS, PL, YZ
Investigation: NH, CVPD, JA
Visualization: NH
Supervision: JXC
Writing—original draft: NH
Writing—review & editing: all authors

**Competing interests:** The authors declare that they have no competing interests.

**Data and materials availability:** All data needed to evaluate the conclusions in the paper are present in the paper and/or the Supplementary Materials. Correspondence and requests for materials should be addressed to JXC.